\documentclass[rnote]{aa}  

\usepackage{graphicx}
\usepackage{txfonts}
\begin{document}

   \title{Pre-outburst observations of Nova Del 2013 from Pan-STARRS\,1}

   \author{N.R. Deacon
          \inst{1}\fnmsep\thanks{Email: deacon@mpia.de}
          \and
          D.W. Hoard\inst{1}\thanks{Visiting Scientist, MPIA}
           \and
          E.A. Magnier \inst{2}
          \and Y.S. Jadhav \inst{1,3}
          \and M. Huber \inst{2} 
          \and K.C. Chambers \inst{2}
          \and H. Flewelling \inst{2}
          \and K.W. Hodapp \inst{4}
          \and N. Kaiser \inst{2}
          \and R.P. Kudritzki \inst{2}
          \and N. Metcalfe \inst{5}
          \and C. Waters \inst{2}
          }
   \institute{Max Planck Institute for Astronomy, K\"onigstuhl 17, Heidelberg, 69117, Germany
   \and
   Institute for Astronomy, University of Hawai`i, 2680 Woodlawn Drive, HI 96822, USA  
   \and
  Dept. of Physics \& Astronomy, College of Arts \& Sciences, Clippinger Labs 251B, Athens, OH 45701, USA
   \and
  Department of Physics, University of Durham, South Road, Durham DH1 3LE, UK 
   \and
   Institute for Astronomy, University of Hawai`i, 640 N. Aohoku Place, Hilo, HI 96720, USA
             }
   \date{Received February 18, 2014; accepted February 23, 2014}

 
  \abstract
   {Nova Delphini 2013 was identified on the 14th of August 2013 and eventually rose to be a naked eye object.}
   {We sought  to study the behaviour of the object in the run-up to outburst and to compare it to the pre-outburst photometric characteristics of other novae.}
   {We searched the Pan-STARRS\,1 datastore to identify pre-outburst photometry of Nova Del 2013 and identified twenty-four observations in the 1.2 years before outburst.}
   {The progenitor of Nova Delphini showed variability of a few tenths of a magnitude but did not brighten significantly in comparison with archival plate photometry. We also found that the object did not vary significantly on the approximately half hour timescale between pairs of Pan-STARRS\,1 observations.}
   {}


   \maketitle
%

\section{Introduction}
Nova Delphini 2013 (=V339 Del) was first identified as a bright 6.8 magnitude source by Koichi Itagaki on the 14th of August 2013 and announced in AAVSO Alert Notice 489 \citep{Waagen2013}. Since then this object has been extensively monitored using radio continuum \citep{Roy2013}, near-infrared \citep{Stringfellow2013}, optical \citep{Darnley2013,Tarasova2013a}, X-ray (with \citealt{Page2013} first detecting emission 36 days after discovery) and gamma-ray \citep{Hays2013} observations. Initially the nova exhibited strong P~Cygni profiles in the Balmer and He~I lines \citep{Tarasova2013a}, these profiles disappeared from the hydrogen lines on the 19th of August \citep{Darnley2013} but remained for other species (He~I, O~I, Fe~II). 

The progenitor of this nova has been identified as USNO-B1.0~1107-0509795 which had an unfiltered magnitude of 17.1 on the 14th of May 2013 \citep{Denisenko2013}. \cite{Wren2013} report a pre-discovery detection as the nova was brightening. Data on this object prior to outburst remain relatively scarce. Denisenko reports a counterpart in the UV from GALEX while there are a handful of detections in POSS photographic plate catalogues. \cite{Tang2013} report three pairs of H$\alpha$ on/off measurements but find no strong evidence that the progenitor is an H$\alpha$ emitter. \cite{Munari2013} identified the progenitor in Asiago photographic plate data, finding 12 detections and 13 upper limits on the object's brightness. Over the course of three years of observations from 1979 to 1982, the progenitor showed variation in the $B$-band with a total amplitude of 0.9\,mag. \cite{Munari2013} also stacked three observations taken by the APASS survey \footnote{http://www.aavso.org/apass} in April 2012 to provide additional $B$ and $V$ photometry.

\subsection{Pre-outburst observations of novae}
\cite{Robinson1975} presented a landmark study of pre- and post- outburst data for novae from archival photographic plates. Out of 18 targets with both pre- and post-outburst magnitude measurements, he identified only one object (BT Monocerotis) with significantly different brightnesses.  Out of 12 targets with well-sampled pre-outburst light curves, five showed an apparent increase in luminosity in the few years before outburst, and two showed significantly increased variability in the lead-up to their outbursts. 

These claims were later re-examined by \cite{Collazzi2009}, who combined subsequent remeasurements of plate data both from their own work and from other sources (whereas \citealt{Robinson1975} relied on the available pool of published photometry from the plates, which can suffer from systematic calibration errors up to $\gtrsim1$~mag). They concluded that BT Mon actually showed no significant change between its pre- and post-outburst brightness and did not reproduce four of the five claims of pre-outburst brightening. They confirmed that the one remaining object from the \cite{Robinson1975} pre-outburst brightening sample (V533 Herculis) did, in fact, increase in brightness by $\sim1.3$~mag in the $\sim1.5$~years leading up to outburst.   \cite{Collazzi2009} identified an additional nova (V1500 Cygni) that showed pre-outburst brightening, rising nearly 7~mag in the month before its outburst\footnote{Although post-outburst behavior is not explicitly the topic of this paper, we note that V1500 Cygni was also one of the five (out of 30) novae that  \cite{Collazzi2009} found to have brightened by more than a factor of 10 after outburst, while the remaining majority of novae had essentially no change in quiescent brightness across the outburst.}.  In addition,  \cite{Collazzi2009} found no case for a class of novae with a significant increase in variability leading up to the outburst, noting that one such object claimed by \cite{Robinson1975} (RR Telescopii) has since been found to be a symbiotic star and that the other (V446 Herculis) actually did not exhibit any change in its variability behavior before and after outburst. They also note that changes in variability amplitude in general are characteristics of novae in quiescence and may not be associated with the outburst -- they conclude that none of the novae they examined showed a significant change in variability that could be linked to the outburst. It should be noted that \cite{Collazzi2009} also found also found a pre-outburst dip in the recurrent nova T CrB. Recent additional archival photometry unearthed by \cite{Schaefer2014} indicates that this object was $\sim1\,mag$ brighter in the 8 years leading up to this brief pre-erruption dip. \cite{Schaefer2013} also report a brief, temporary brightening of $\sim$1\,mag of  another recurrent nova, T Pyx, prior to outburst.

The total brightness of cataclysmic variables (CVs) with accretion disks -- of which classical novae are a subset -- are dominated by accretion-generated luminosity over a wide range of the electromagnetic spectrum (from the near-IR to the far-UV).  As such, variability in the long-term\footnote{``Long-term'', in this case, is relative since it might easily be measured in units of the CV's orbital period.  At typical values of $\sim1$--$10$~hr, even many orbital cycles span an interval that is, in an absolute sense, not very long.  CVs, of course, famously display a wide range of short timescale, low amplitude variability - ``flickering'' -- but this is likely related to the effects of viscous turbulence in the disk and/or accretion stream and not to actual changes in the mean mass transfer rate from the donor star \citep{Scaringi2013}.}. mean brightness is directly proportional to changes in the mass transfer rate from the low mass donor star, through the disk, and onto the white dwarf (WD) primary star.  The origin of a classical nova outburst is a thermonuclear runaway in the thin layer of accreted hydrogen-rich matter on the surface of the WD.  The total mass in this layer required to trigger the runaway is a few $\times10^{-3}$~M$_{\odot}$ for a WD with $M_{\rm WD}=0.6$~M$_{\odot}$ (i.e., the average WD mass) to a few $\times10^{-5}$~M$_{\odot}$ for WDs approaching the Chandrasekhar mass (see \citealt{Warner2003} and references therein).  Even under the optimistic assumption of a persistent high CV mass transfer rate of $\sim10^{-9}$~M$_{\odot}$~yr$^{-1}$, the timescale to grow this accreted envelope to the critical mass is $\gtrsim10^{4}$--$10^{6}$~yr. 

This is the root of the reason that \cite{Collazzi2009} found Robinson's conclusion that almost 50\% of novae show a pre-outburst brightness increase to be ``unsettling''.  They discuss in some detail the arguments against a corresponding pre-outburst increase in the mass transfer rate being the ``last straw'' that triggers the nova outburst.  In summary, (a) the amount of extra mass accumulated during the rise in brightness is miniscule compared to the total mass required to trigger the outburst, and (b) the energy flux carried by the extra accreted matter produces an insignificant increase in the temperature at the base of the accumulated layer on the WD.  The result of  \cite{Collazzi2009}, two out of 22 novae ($<10$\%) displaying a pre-outburst rise in brightness, might well be less unsettling, but it is still problematic:\ as they note, the close proximity to the nova outbursts in both V533 Her and V1500 Cyg suggests that the pre-outburst brightness increases are somehow causally connected to the outbursts, although they cannot have directly triggered the outbursts.  

In the majority of classical novae, it appears, the outburst should arrive as a surprise, without previous announcement in the light curve.  It is the currently unexplained exceptions to this situation, like V533 Her and V1500 Cyg, that might offer clues that will help to further illuminate the nova outburst mechanism.  Similarly, although both \cite{Robinson1975} and  \cite{Collazzi2009} agree that the majority of novae have the same pre- and post-outburst mean quiescent brightness (implying that the outburst does not significantly alter the mass transfer process in the CV), the latter found that $\sim15$\% of novae {\em do} show a post-outburst increase in brightness of $>2.5$~mag.  Again, the few exceptions to the majority behavior might offer the best insight into the nova outburst mechanism.  Obtaining post-outburst data on novae is easy, but for pre-outburst data, we must typically rely on serendipitous observations.  In this paper we describe pre-outburst photometry from Pan-STARRS\,1 for Nova Del 2013.

\section{Observations}
\subsection{Pan-STARRS\,1 observations}
Pan-STARRS\,1 is a 1.8~m high-etendue survey telescope situated on Haleakala on Maui in the Hawaiian Islands \citep{Kaiser2002}. Approximately 50\% of the time on this telescope is devoted to the 3$\pi$ survey which is covering the sky north of $-30^{\circ}$ in five filters ($g_{P1}$, $r_{P1}$, $i_{P1}$, $z_{P1}$ and $y_{P1}$ with effective wavelengths of 481, 617, 752, 866 and 962\,nm respectively; \citealt{Tonry2012}) with six pairs of observations per filter in each area of sky over the course of the survey. The survey photometric reference system and calibration are described by \cite{Magnier2013} and \cite{Schlafly2012} respectively. The 3$\pi$ survey strategy is designed to   accommodate various science and observing constraints. Observations in $z_{P1}$ and $y_{P1}$ are taken 3--4 months before opposition to maximise the parallax factor for late-type objects. The two pairs of observations in each of $g_{P1}$, $r_{P1}$ and $i_{P1}$ are usually taken in one lunation close to opposition. Observations in the same filter are taken in pairs separated by approximately 25 minutes. These pairs are used for asteroid detection but are also useful in the study of cataclysmic variables, spanning a significant portion of the typical short period variability timescales for such objects. Hence variability between the observations in a pair is a characteristic of CVs (Jadhav et al., in prep., \citealt{Jadhav2014}).

We extracted all data for Nova Delphini 2013 from the Pan-STARRS\,1 data store. In order to plot only points with reliable photometry we followed the suggestion of \cite{Morganson2012} and only selected observations with a PSF quality factor (the fraction of pixels from the object PSF which were not saturated and did not fall on bad pixels or chip-gaps) greater than 0.8. These are listed in Table~\ref{ps1phot}.

\begin{table}
\caption{Photometry of Nova Del 2013 from Pan-STARRS\,1. The psf\_qf parameter describes the fraction of pixels from the object PSF which were not saturated and did not fall on bad pixels or chip-gaps. }             
\label{ps1phot}      
\centering                          
\begin{tabular}{l c c c}        
\hline\hline                 
magnitude(AB)&psf\_qf&filter&JD\\    
\hline                        
17.569$\pm$0.029&0.998&$i_{P1}$&2456138.973\\
17.583$\pm$0.033&0.999&$i_{P1}$&2456138.984\\
16.914$\pm$0.029&0.998&$g_{P1}$&2456149.811\\
16.940$\pm$0.024&0.998&$g_{P1}$&2456149.821\\
16.986$\pm$0.041&0.999&$r_{P1}$&2456150.783\\
16.999$\pm$0.047&0.998&$r_{P1}$&2456150.794\\
17.733$\pm$0.027&0.998&$z_{P1}$&2456228.763\\
17.770$\pm$0.029&0.999&$z_{P1}$&2456228.774\\
17.576$\pm$0.027&0.999&$z_{P1}$&2456406.089\\
17.603$\pm$0.022&0.999&$z_{P1}$&2456406.101\\
17.297$\pm$0.024&0.998&$y_{P1}$&2456409.103\\
17.331$\pm$0.022&0.999&$y_{P1}$&2456409.115\\
17.474$\pm$0.029&0.999&$y_{P1}$&2456464.123\\
17.349$\pm$0.022&0.997&$y_{P1}$&2456466.984\\
18.325$\pm$0.022&0.998&$r_{P1}$&2456475.070\\
18.278$\pm$0.020&0.998&$r_{P1}$&2456475.080\\
17.206$\pm$0.031&0.998&$i_{P1}$&2456477.076\\
17.236$\pm$0.046&0.997&$i_{P1}$&2456477.087\\
17.040$\pm$0.042&0.999&$g_{P1}$&2456481.046\\
17.020$\pm$0.026&0.997&$g_{P1}$&2456481.056\\
17.414$\pm$0.030&0.999&$g_{P1}$&2456486.869\\
17.451$\pm$0.034&0.998&$g_{P1}$&2456486.879\\
17.244$\pm$0.031&0.998&$i_{P1}$&2456499.872\\
17.287$\pm$0.026&0.998&$i_{P1}$&2456499.879\\
12.831$\pm$0.020\tablefootmark{\dag}&0.886&$y_{P1}$&2456565.704\\
\hline                                   
\end{tabular}
\tablefoot{\dag This post-outburst datapoint has a slightly low psf\_qf and is close to the saturation limit in the $y_{P1}$ band. Hence it should be considered potentially unreliable.}
\end{table}

\subsection{Other photometry}
\begin{figure*}
   \resizebox{\hsize}{!}
            {\includegraphics{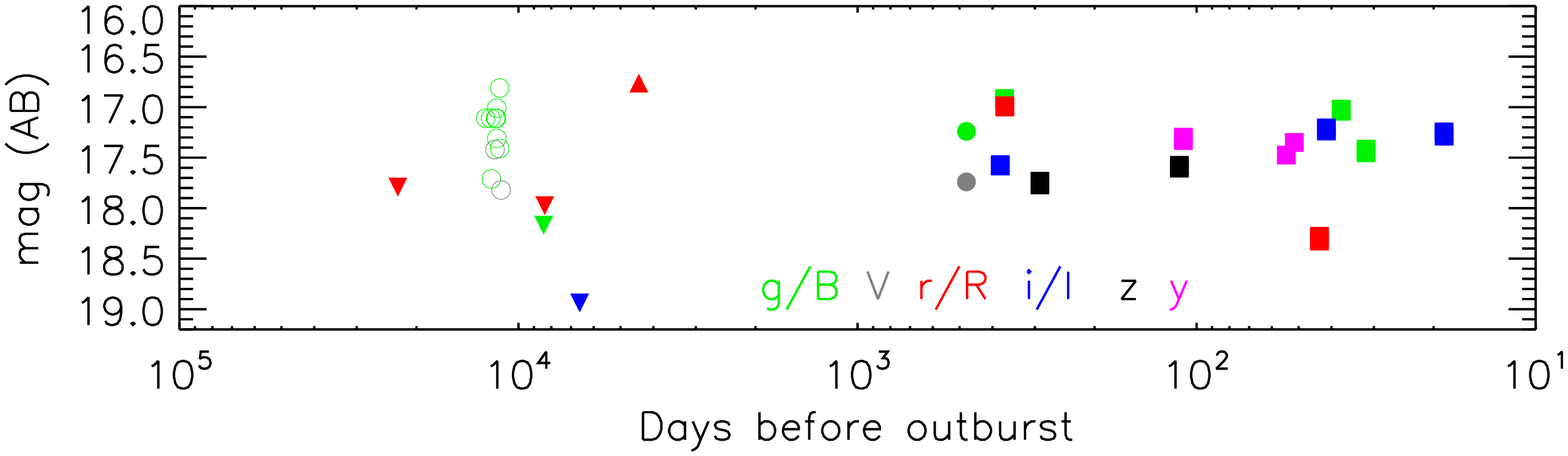}}
      \caption{Pre-outburst photometry for Nova Del 2013. Square points are from Pan-STARRS\,1, downward facing triangles from plate data, upward facing triangles from the Carlsberg Meridian catalogue, filled circles are APASS data and points marked with open circles come from Asiago photographic plates.}
         \label{photplot}
   \end{figure*}

We extracted observations from other sources by searching various astronomical archives (see Table~\ref{otherphot}). Our main source comes from the SuperCOSMOS Sky Survey \citep{Hambly2001} scans on Palomar POSS plates.  While there are no errors quoted for the SuperCOSMOS fields, \cite{Hambly2001a} quote approximate errors for UKST filters as a function of magnitude. Hence for the POSS-II $R_F$ point the approximate error will be 0.14 mag, 0.34 mag in POSS-II $B_J$ and 0.17 mag. in POSS-II $I_N$. These datapoints are in a system on the Vega scale; for our purposes we converted these using the Vega-AB corrections of \cite{Blanton2007}. Note that these transformations are not for exactly the same filter set as the POSS plates but are the closest available. We applied the same process to the APASS and Asiago photometry from \cite{Munari2013}. The Asiago data do not have formal errors for each observation but \cite{Munari2013} quote an error of 0.1\,mag for all datapoints. All of the available broadband pre-outburst data for Nova Del 2013 are shown in Figure~\ref{photplot}.
\begin{table}
\caption{Photometry of Nova Del 2013 from other sources. The POSS plate photometry is on the Vega scale and comes from the SuperCOSMOS Sky Survey \citep{Hambly2001}. The Carlsberg Meridian \citep{CMC2006} point is on the AB scale and the Asiago and APASS measurements are on the Vega scale.}             
\label{otherphot}      
\centering                          
\begin{tabular}{l c c}        
\hline\hline                 
magnitude&filter&JD\\    
\hline                        
17.58&POSS-I $E$&2433835.923\\
17.2&Asiago $B$&2444016.506\\
17.2&Asiago $B$&2444436.502\\
17.8&Asiago $B$&2444526.399\\
17.4&Asiago $V$&2444812.455\\
17.2&Asiago $B$&2444840.491\\
17.2&Asiago $B$&2444854.431\\
17.2&Asiago $B$&2444901.382\\
17.1&Asiago $B$&2444925.258\\
17.4&Asiago $B$&2444931.316\\
17.5&Asiago $B$&2445148.554\\
16.9&Asiago $B$&2445166.542\\
17.8&Asiago $V$&2445263.365\\
18.26&POSS-II $B_J$&2448091.863\\
17.77&POSS-II $R_F$&2448150.706\\
18.49&POSS-II $I_N$&2449917.867\\
16.76\tablefootmark{\dag}&CMC $r'$&2452102\\
17.33$\pm$0.09\tablefootmark{\ddag}&APASS $B$&2456040.833\\
17.06$\pm$0.10\tablefootmark{\ddag}&APASS $V$&2456040.833\\
\hline                                   
\end{tabular}
\tablefoot{\dag The Carlsberg Meridian catalogue lists the two datapoints for
  the object as being non-photometric. Additionally the epoch and magnitude
  are means of the two observations.}
\tablefoot{\ddag The APASS measurements result from \cite{Munari2013}
  stacking three observations from the nights of the 21st, 24th and 25th of
  April 2012. Here we assign APASS observations the mean epoch of these three dates.}
\end{table}

\section{Discussion}
Figure~\ref{photplot} shows that the progenitor of Nova Del 2013 was variable in the lead-up to its outburst. Table~\ref{phot_var} quantifies the variability in each Pan-STARRS\,1 band. There is an apparent trend of increasing variability with decreasing wavelength as expected from flickering in CVs \citep{Bruch1992}; however, it is impossible to state  with certainty if this trend is real or the result of sparse sampling. We also note that none of the pre-outburst pairs of observations show differences between their components of more than 2$\sigma$. This indicates that the progenitor was not significantly variable on timescales of approximately half an hour. Additionally we saw no dwarf-nova-like outbursts of a magnitude or more. While the Pan-STARRS\,1 data are marginally brighter than the POSS photographic data (excluding the $I_N$ observation which is close to the sensitivity limit for the $I_N$ plates) there is no evidence for a significant increase in brightness from less than a month to several years before the outburst, as was observed for V1500 Cyg and V533~Her, respectively \citep{Collazzi2009}.  Ignoring the influence of different bandpasses and considering the data in Figure 1 {\it en masse}, the recent Pan-STARRS 1 photometry of Nova Del 2013 is possibly $\sim0.5$~mag brighter than the much earlier SuperCOSMOS archival observations (but not the APASS, Carlsberg or Asiago data).  Even if this is a real feature and not, for example, an artifact resulting from the comparison of historic photographic photometry with modern CCD photometry, then it is within the limit of the expected level of ``normal'' variability based on observations of many nova light curves away from outburst \citep{Collazzi2009}. Once Nova Del 2013 fades back to quiescence, it will be illuminating to compare its post-outburst behaviour to the pre-outburst data measured in this paper.

\begin{table}
\caption{The variability of Nova Del 2013 in Pan-STARRS\,1 data taken in the 1.2 years running up to the outburst. }             
\label{phot_var}      
\centering                          
\begin{tabular}{l c c c}        
\hline                
filter&mean&$\sigma$&$n_{meas}$\\    
\hline 
$g_{P1}$&17.130&0.219&6\\
$r_{P1}$&17.647&0.655&4\\
$i_{P1}$&17.354&0.159&6\\
$z_{P1}$&16.671&0.082&4\\
$y_{P1}$&17.363&0.067&4\\
\hline                        

\hline                                   
\end{tabular}
\end{table}
\section{Conclusions}
We find that prior to its eruption, Nova Del 2013 did not show a significant change in photometric behaviour. Similar to most of the novae studied by \cite{Collazzi2009}, the progenitor did not show a significant increase in brightness in the few years prior to outburst. The lack of significant variability between the components of Pan-STARRS\,1 observation pairs indicates that the progenitor was not significantly variable on the timescale of approximately half an hour. 

During the next several years, a similar amount of pre-outburst photometry will be available for the majority of northern novae from the ongoing Pan-STARRS project.  A number of similar ongoing and planned wide-field, temporal surveys will provide additional multi-epoch photometric coverage of nova progenitors over, collectively, a large fraction of the entire sky in both hemispheres (e.g., the Catalina Real-time Transient Survey, \citealt{Drake2012}; the Palomar Transient Factory, \citealt{Rau2009}; SkyMapper, \citealt{Keller2013}).  Moving into the next decade, the more densely sampled observing cadence of the Large Synoptic Survey Telescope (LSST; \citealt{Ivezic2008}) will provide even more data that can be used to characterize the behavior of novae during the (potentially long-term) run-up to outburst.  The arduous process of combing through archival plates in the hopes of serendipitously locating nova progenitors, and then processing those images to provide usable photometric data, will be increasingly supplanted by an ever-growing spatial and temporal database of accurate, homogeneous, digital photometry extending to very faint objects.  This will finally allow us to systematically assess the presence and characteristics of photometric changes displayed by classical novae in relation to their outbursts, providing crucial input and constraints to modelling the physics behind these accretion-induced thermonuclear events.

\begin{acknowledgements}
 The Pan-STARRS1 Surveys (PS1) have been made possible through contributions of the Institute for Astronomy, the University of Hawaii, the Pan-STARRS Project Office, the Max-Planck Society and its participating institutes, the Max Planck Institute for Astronomy, Heidelberg and the Max Planck Institute for Extraterrestrial Physics, Garching, The Johns Hopkins University, Durham University, the University of Edinburgh, Queen's University Belfast, the Harvard-Smithsonian Center for Astrophysics, the Las Cumbres Observatory Global Telescope Network Incorporated, the National Central University of Taiwan, the Space Telescope Science Institute, the National Aeronautics and Space Administration under Grant No. NNX08AR22G issued through the Planetary Science Division of the NASA Science Mission Directorate, the National Science Foundation under Grant No. AST-1238877, the University of Maryland, and Eotvos Lorand University (ELTE). This research has made use of data obtained from the SuperCOSMOS Sky Survey, prepared and hosted by the Wide Field Astronomy Unit, Institute for Astronomy, University of Edinburgh, which is funded by the UK Science and Technology Facilities Council. This research has made use of the APASS database, located at the AAVSO web site. Funding for APASS has been provided by the Robert Martin Ayers Sciences Fund. We thank our anonymous reviewer for a helpful and extremely swift referee report.
\end{acknowledgements}


\bibliographystyle{aa}
\bibliography{ndeacon}{}
\end{document}